\documentclass[twoside]{dis07}
\usepackage[latin1]{inputenc}
\usepackage[dvips]{graphicx,epsfig,color}
\usepackage{wrapfig,rotating}
\usepackage{amssymb,amsmath,array}

\begin{document}
\title{Review of Beauty Production at HERA and Elsewhere}

\author{A.Geiser
%
\vspace{.3cm}\\
%
DESY Hamburg, Germany
}
%
%
%

\maketitle

\begin{abstract}
Experimental results on beauty production at HERA are reviewed 
in the context of similar measurements at other colliders.
As a result of a phenomenological study of the 
QCD scale dependence of many different NLO and NNLO predictions,
a modification of the ``default'' scale choice is advocated.
Experimental constraints on the photon-quark coupling are also 
investigated. \cite{slides}   
\end{abstract}

\section{Introduction}

%
%
%
%
\begin{wrapfigure}{r}{0.25\columnwidth}
\centering
\centerline{\includegraphics[width=0.20\columnwidth]{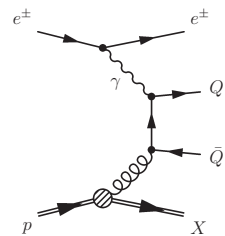}}
\caption{\small Feynman graph for the production of a heavy quark pair via the 
boson-gluon-fusion (BGF) process.}
\label{fig:BGF}
\end{wrapfigure}
%
%
Beauty production at HERA (Fig. 1) is an important tool to investigate our 
present understanding of the theory of Quantum Chromo-Dynamics (QCD).
On one hand, the large $b$ quark mass, taken as a hard scale, ensures that 
the cross sections are
always perturbatively calculable. On the other hand, the simultaneous 
presence of competing hard scales, such as the transverse momentum ($p_T$)
of the heavy quark, or the virtuality of the exchanged photon ($Q^2$), induces 
additional theoretical uncertainties due to terms in the perturbative 
expansion which depend logarithmically on the ratio of these scales.   
The comparison of the measured cross sections with 
theory predictions is therefore particularly sensitive to the way the 
perturbative expansion is made, and can therefore potentially discriminate
how adequate a particular QCD scheme is for the decription of the cross section
in question.
This can also yield insights for other QCD processes at HERA, and for related
processes at other colliders, including future measurements at the LHC.

Since beauty in Deep Inelastic Scattering (DIS) is covered
elsewhere \cite{Kahle}, this contribution will concentrate on
the photoproduction case ($Q^2 < 1$ GeV$^2$), in which the photon is 
quasi-real.
For beauty photoproduction at HERA, possible theoretical schemes include

\noindent $\bullet$ 
The leading order plus parton shower approach, where leading order (LO) 
QCD matrix elements are complemented by parton showers, usually using the 
DGLAP \cite{DGLAP} parton evolution equations. This approach is implemented in 
many Monte Carlo models, and mostly used for
the purpose of acceptance corrections.

\noindent $\bullet$ The kt-factorization approach \cite{CCFM}, which can alternatively be 
used for parton showering, combined with the use of generalized parton 
density functions. 

\noindent $\bullet$
The next-to-leading order (NLO) massive approach \cite{Frixione:1995qc}. 
In this approach,
the heavy quark mass is fully accounted for, and heavy quarks are therefore
always produced dynamically in the matrix element, as illustrated by Fig. 1.
Alternative LO processes, such as flavour excitation in the photon or the 
proton, are treated as next-to-leading order corrections to this BGF process.
Processes in which the photon acts as a hadron-like source of light quarks 
or gluons are also included, but make only a small contribution.
This approach is expected to work best when all relevant hard scales, 
e.g. $p_T$, are of order $m_b$.

\noindent $\bullet$ 
For $p_T \gg m_b$, large $\log p_T/m_b$ terms could in principle spoil
the reliability of the predictions. In this case, it might be preferable to
switch to a so-called massless scheme, in which the $b$ mass is neglected
kinematically. 
The potentially large logarithms can then be resummed to all orders (next-to
leading log or NLL resummation).
Since such an approach is obviously not applicable when $p_T \sim m_b$,
schemes have been designed which make a continous transition between 
the fixed order (FO) massive, and the NLL massless scheme. This is often
referred to as the FONLL scheme
\cite{Cacciari:2001td}.  

%
%
%
%
\begin{wrapfigure}{r}{0.55\columnwidth}
\centering
\vspace{-1.0cm}
\includegraphics[width=0.6\columnwidth]{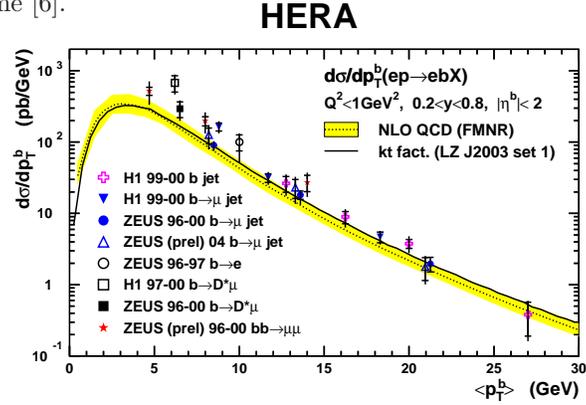}
\vspace{-1.0cm}
\caption{\small
Beauty production cross section
measurements in photoproduction at HERA
as function of the transverse momentum of the $b$ quark,
compared to QCD predictions.
}
\label{fig:ptb}
\end{wrapfigure}
On the experimental side, several different methods are used to tag the 
beauty final state.  The $b$ quark can decay semileptonically into
a muon or electron, which can be identified in the detector. The large
momentum of the lepton transverse to the direction of the $b$-initiated jet,
due to the sizeable $b$ mass, can be used to discriminate against semileptonic
charm decays or misidentified light flavour events.  
The finite lifetime of the $B$ hadrons can lead to 
a measureable offset of the decay vertex with respect to the primary vertex
of the event, which also leads to a significant impact parameter of the 
resulting secondary tracks. Finally, a lepton tag can e.g. be combined 
with a lifetime tag, with a second lepton tag, or with a $D^*$ meson from 
a $b$ decay. 

%
%
\begin{wrapfigure}{t}{0.45\columnwidth}
\vspace{-0,5cm}
\includegraphics[width=0.45\columnwidth]{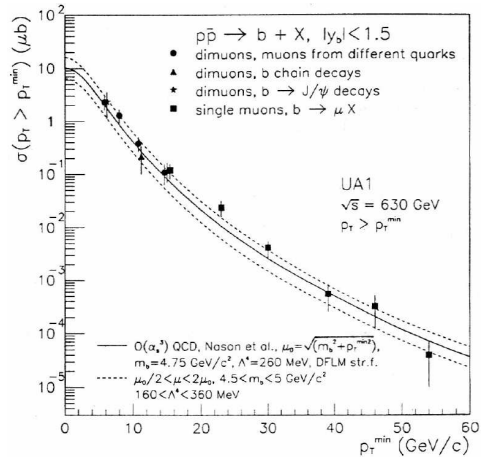}
\caption{\small
Integrated beauty production cross section at the $Sp\bar pS$
as function of the minimum $p_T$ of the $b$ quark,
compared to NLO QCD predictions.
}
\label{fig:ptbUA1}
\end{wrapfigure}
Fig. \ref{fig:ptb} shows a compilation of all recent HERA measurements of $b$ 
photoproduction \cite{bmeasurements}. Reasonable agreement is found with
both the fixed order NLO QCD prediction \cite{Frixione:1995qc}, 
and with a prediction 
based on kt factorization \cite{LZ}. However, the data tend to lie somewhat 
above the central prediction in both cases.

A longstanding apparent discrepancy between data and theory in $b$ production
at the Tevatron was resolved by combining a more careful
consideration of $B$ fragmentation and decay parameters with an FONLL-based
prediction \cite{CaccNas}.
This raises the question whether an FONLL predicton, which does not yet 
exist for $b$ production at HERA, would yield an improved agreement. 

\begin{figure}[hb]
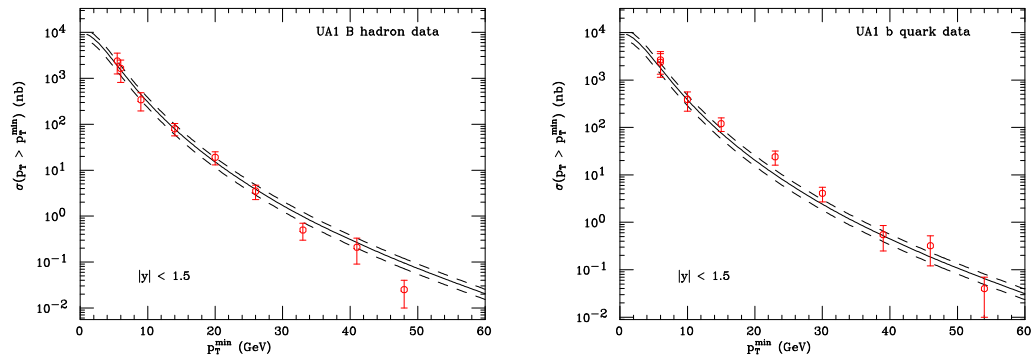

\includegraphics[width=0.46\columnwidth]{geiser_achim.fig4a.eps}
\hspace{0.5cm}
\includegraphics[width=0.46\columnwidth]{geiser_achim.fig4b.eps}
\caption{\small
Integrated beauty production cross section at the $Sp\bar pS$
as function of the minimum $p_T$ of the $B$ hadron (left)
and $b$ quark (right),
compared to preliminary FONLL QCD predictions \cite{Caccpriv}.
}
\label{fig:UA1FONLL}
\end{figure}

For this purpose, consider $b$ production at the $Sp\bar pS$ collider, which
had an effective parton-parton center-of-mass energy very similar to that of 
HERA.
Fig. \ref{fig:ptbUA1} shows the measured $b$ quark cross section \cite{UA1}
compared to the original NLO calculation \cite{MNR}. Good agreement was 
observed at a time when the Tevatron experiments were starting to claim
a discrepancy.
Fig. \ref{fig:UA1FONLL} shows the same original data \cite{UA1} compared to 
the more recent FONLL calculation at $b$ quark and $B$ hadron level 
\cite{CaccNas,Caccpriv}, with identical parameters as those used 
for the Tevatron.
Good agreement is observed, also at muon level \cite{Caccpriv}, even though
NLO predictions at $B$ hadron and muon level were not available when the 
measurements were made.
This indicates that the $B$ fragmentation and decay spectra, which had
been studied carefully \cite{UA1,AG}, were treated consistently
in these measurements.   
Furthermore, the NLO and FONLL predictions agree very well with each other,
indicating that the large logs mentioned above do not yet play a significant 
role in this $p_T$ range (similar to the one at HERA).
This can also be seen in charm production at HERA \cite{zeus-ichep02-786} for which 
an FONLL prediction exists.

%
%
In conclusion, an FONLL prediction for beauty production at HERA would be 
useful, but is not expected to significantly alter the data/theory comparison.

The dominant contribution to the theoretical error band of Fig. 1 is 
the variation of the renormalization/factorization scale by a factor 2 
around the default scale $\mu_0 = \sqrt{m_b^2 + p_T^2}$.
Such a variation is intended to reflect the uncertainty due to uncalculated
higher orders.
It might therefore be useful to reconsider this scale choice.

Ideally, in a QCD calculation to all orders, the result of the perturbatve
expansion does not depend on the choice of these scales\footnote{As is 
common practice, we will not distinguish between the 
factorization and renormalization scales in the following, and set both to be 
equal. 
A separate optimization of the two scales, which should be 
done in principle, will be left for future consideration.}.
In practice, a dependence arises from the truncation of the 
perturbative series. Since this is an artefact of the truncation, rather than
a physical effect, the optimal scale can not be ``measured'' from the data.
It must thus be obtained phenomenologically.  

Traditionally, there have been several options to choose the ``optimal'' scale,
e.g.

\noindent $\bullet$ 
The ``natural'' scale of the process. This is usually taken to be the 
transverse 
energy ($E_T$) of the jet for jet measurements, the mass $m$ of a heavy 
particle for the total production cross section of this particle, or the 
combination $\sqrt{m^2+p_T^2}$ for differential cross sections of such a 
particle.
Often, this is the only option considered.
The choice of this natural scale is based on common sense, and on 
the hope that this will minimize the occurrance of large logs
of the kind described above, for the central hard process.
However, higher order subprocesses such as additional gluon radiation often
occur at significantly smaller scales, such that this choice might not
always be optimal.

\noindent $\bullet$ 
The principle of fastest apparent convergence (FAC) \cite{FAC}.
The only way to reliably evaluate uncalculated higher orders
is to actually do the higher order calculation. Unfortunately, this is often 
not possible. Instead, one could hope that a scale choice which 
makes the 
leading order prediction identical to the next-to-leading order one would 
also minimize the NNLO corrections. Since it can not be proven, 
this principle, which can be found in many 
QCD textbooks, has not been used very much recently. However, recent
actual NNLO calculations might indicate that it works phenomenologically 
after all (see below).

\noindent $\bullet$ 
The principle of minimal sensitivity (PMS) \cite{PMS}. The idea is 
that when the derivative of the cross section with respect to the NLO scale 
variation vanishes, the NNLO crrections will presumably also be small.
Again, there is no proof that this textbook principle should work, but
actual NNLO calculations might indicate that it does (see below).    
 
\begin{figure}[ht]
\includegraphics[width=0.5\columnwidth]{geiser_achim.fig5a.eps}
\hspace{1 cm}
\includegraphics[width=0.31\columnwidth]{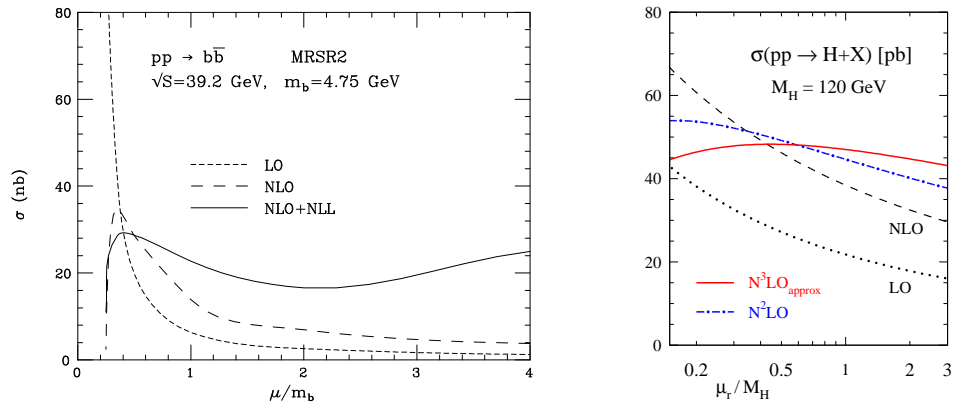}
\caption{\small
Scale dependence of the total cross section for beauty production at HERA-B 
\cite{HERABpred} (left) and for Higgs production at LHC \cite{HiggsLHC} for two different masses (right).
}
\label{fig:scalevar}
\end{figure}

To illustrate these principles, consider two examples.
First, the prediction for the total cross section for beauty production at 
HERA-B \cite{HERABpred} (Fig. \ref{fig:scalevar}). 
The natural scale for this case is the $b$ quark mass, $\mu_0 = m_b$,
and all scales are expressed as a fraction of this reference scale. 
Inspecting Fig. \ref{fig:scalevar}, one finds that both the PMS and FAC 
principles, applied to the NLO prediction and to the comparison with LO
(NLO stability), would yield an optimal scale of about half the natural scale.
The same conclusion would be obtained by using the NLO+NLL prediction, 
including resummation, and comparing it to either the LO or the NLO 
prediction (NLO+NLL stability).         
 
%
%
Second, the prediction for Higgs production at the LHC \cite{HiggsLHC} (Fig. \ref{fig:scalevar}).
The reference scale is now the Higgs mass ($\mu_0 = m_H$).
However, inspecting the behaviour of the LO and NLO predictions, neither the
FAC nor the PMS principle would yield a useful result in this case, since the
two predictions do not cross, and the NLO prediction does not have a maximaum
or minimum.
This situation occurs rather frequently, and is also true for $b$ production 
at HERA. 
Fortunately, in the case of Higgs production, the NNLO and even NNNLO 
predictions have actually been calculated (Fig \ref{fig:scalevar}).
Applying the FAC and PMS prescriptions to these instead (NNLO stability), again
a scale significantly lower than the default scale would be favoured.
This might indicate that choosing a scale which is smaller than the default 
one makes sense even if the FAC and PMS principles do not yield useful
values at NLO.

Beyond these examples, a more general study is needed 
to phenomenologically validate this approach.
To avoid additional complications arising from a multiple scale problem 
caused by e.g. the scale $Q^2$ at HERA or the scale $M_Z$ at LEP, the study 
was 
limited to cross sections for photoproduction at HERA, or hadroproduction at
fixed target energies, the Tevatron, and LHC.
The somewhat arbitrary selection of processes includes 
beauty production at the $Sp\bar pS$ \cite{UA1alpha,Frixheavy}, the Tevatron \cite{Frixheavy},
and HERA-B \cite{HERABpred},
top production at the Tevatron \cite{HERABpred,Frixheavy},
direct photon production at fixed target \cite{photon},  
Z \cite{ZLHC} and Higgs \cite{HiggsLHC} production at the LHC,
jets at HERA \cite{jetPHP} and at the Tevatron \cite{TevaJet}.   
This selection is obviously not complete. However, it is not biased in the 
sense that all processes that were considered were included, and none were 
discarded.

%
%
\begin{wrapfigure}{t}{0.45\columnwidth}
\vspace{-0.5cm}
\includegraphics[width=0.45\columnwidth]{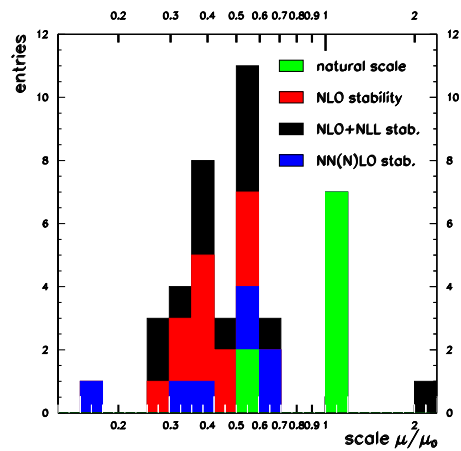}
\vspace{-0.5cm}
\caption{\small
Summary of optimized scales derived as described in the text.
\vspace{-0.5cm}
}
\label{fig:scale}
\end{wrapfigure}
In each case the natural scale as defined above was used as a reference. 
In addition, wherever possible, the optimal scales
from both the FAC and PMS principles, evaluated at NLO (NLO stability), NLO+NLL
(NLO+NLL stability), and/or NNLO/NNNLO (NNLO stability) were evaluated 
separately. Fig. \ref{fig:scale} shows the result of this evaluation.
Each crossing point, maximum, or minimum in Fig. \ref{fig:scalevar} yields one 
entry into this figure, and similarly for all the other processes.
The conclusion is that the FAC and PMS principles tend to favour scales which
are around 25-60\% of the natural scale. Amazingly, this seems to be 
independent of whether these principles are applied at NLO, NLO+NLL, or NNLO 
level. For the jet \cite{TevaJet} or b-jet \cite{Tevabjet} cross sections at the 
Tevatron, it has in part already become customary to use half the natural scale
as the central scale.

Using the natural scale as the default and varying it by a factor two, which
is the choice adopted for most data/theory comparisons, covers only about half
the entries, while the other half lies entirely below this range.
Instead, using half the natural scale as the default and varying it 
by a factor two, thus still including the natural scale in the variation, 
covers about 95\% of all the entries.    

This yields the following conclusions.

\noindent $\bullet$
Obviously, whenever an NNLO calculation is available, it should be used.

\noindent $\bullet$
Whenever possible, a dedicated scale study should be made for each 
process for the kinematic range in question.
Although there is no proof that the FAC and PMS principles should work, 
in practice they seem to give self-consistent and almost universal answers
for processes at fixed target energies, HERA, the Tevatron, and the LHC.

\noindent $\bullet$
In the absence of either of the above, the default scale should be chosen
      to be {\sl half} the natural scale, rather than the natural scale,
      in particular before claiming a discrepancy between data and theory.
      Empirically, this should enhance the chance that the NNLO calculation, 
      when it becomes available, will actually lie within the quoted error 
      band.

\newpage

%
%
%
%
\begin{wrapfigure}{r}{0.55\columnwidth}
\centering
\vspace{-0.5cm}
\includegraphics[width=0.6\columnwidth]{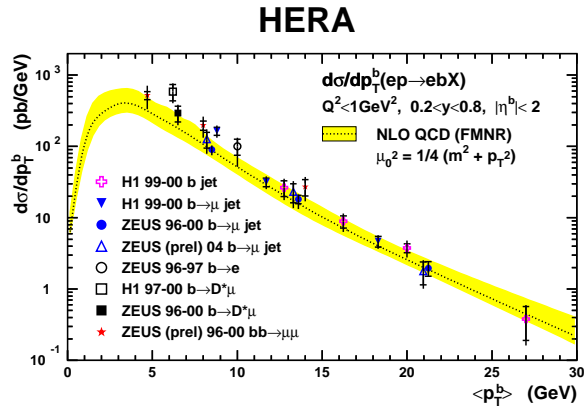}
\vspace{-1.0cm}
\caption{\small
Beauty production cross section
measurements in photoproduction at HERA
as function of the transverse momentum of the $b$ quark,
compared to QCD predictions.
\vspace{-0.3cm}
}
\label{fig:ptbhalf}
\end{wrapfigure}
Now consider the application of the last proposal to actual data/theory 
comparisons.  
Fig.~\ref{fig:ptbhalf} shows the resulting comparison for beauty production
at HERA. Although before the change of the default scale the agreement was 
already quite reasonable, the new choice, based on theoretical/phenomenological
arguments, improves the agreement. 
A similar statement \cite{slides} can qualitatively be made for beauty 
production at the 
$Sp\bar pS$ \cite{UA1}, the Tevatron \cite{Tevab}, and even at 
RHIC \cite{RHIC}, where a discrepancy has been 
claimed. The agreement with charm production at HERA \cite{zeus-ichep02-786} 
as well as 
charm \cite{Tevac} and top \cite{Tevat} 
production at the Tevatron also improves, or at least does not get worse.
The same is true for inclusive jets at HERA, both in DIS \cite{jetDIS} and in 
photoproduction \cite{jetPHP}. In one case \cite{jetscale} half the scale has 
already been used in a published HERA result.

So far, no example is known to the author where the proposed
change of default scale would result in a significant worsening of the 
data/theory agreement in a photo- or hadroproduction cross section.  
Thus, the phenomenologically motivated change seems to be supported by the 
data. It should therefore be considered to make it the default for
future predictions at HERA, the Tevatron, and the LHC. 

The investigations of the photon quark coupling are reported 
elsewhere \cite{slides}.

  
\begin{footnotesize} 
\bibliographystyle{unsrt} 
{\raggedright
\bibliography{geiser_achim}
}
\end{footnotesize} 
\end{document}